\documentclass[sigconf]{acmart}

\AtBeginDocument{%
  }

\setcopyright{acmlicensed}
\copyrightyear{2018}
\acmYear{2018}
\acmDOI{XXXXXXX.XXXXXXX}

\acmConference[WWW Conference]{Make sure to enter the correct
  conference title from your rights confirmation emai}{2023}{Austin,Texas}
\acmISBN{978-1-4503-XXXX-X/18/06}

\begin{document}

\title{Heterogeneous Graph-based Framework with Disentangled Representations Learning for Multi-target Cross Domain Recommendation}


\author{Xiaopeng Liu}
\affiliation{%
  \institution{NetEase Media Technology (Beijing)
Co.,Ltd.}
  \city{Beijing}
  \country{China}}
\email{liuxiaopeng02@corp.netease.com}

\author{Juan Zhang}
\affiliation{%
  \institution{NetEase Media Technology (Beijing)
Co.,Ltd.}
  \city{Beijing}
  \country{China}}
\email{zhangjuan1@corp.netease.com}

\author{Chongqi Ren}
\affiliation{%
  \institution{NetEase Media Technology (Beijing)
Co.,Ltd.}
  \city{Beijing}
  \country{China}}
\email{renchongqi@corp.netease.com}

\author{Shenghui Xu}
\affiliation{%
  \institution{NetEase Media Technology (Beijing)
Co.,Ltd.}
  \city{Beijing}
  \country{China}}
\email{xushenghui@corp.netease.com}

\author{Zhaoming Pan}
\affiliation{%
  \institution{NetEase Media Technology (Beijing)
Co.,Ltd.}
  \city{Beijing}
  \country{China}}
\email{panzhaoming@corp.netease.com}

\author{Zhimin Zhang}
\affiliation{%
  \institution{NetEase Media Technology (Beijing)
Co.,Ltd.}
  \city{Beijing}
  \country{China}}
\email{zhangzhimin@corp.netease.com}


\begin{abstract}
  CDR (Cross-Domain Recommendation), i.e., leveraging information from multiple domains, is a critical solution to data sparsity problem in recommendation system. The majority of previous re search either focused on single-target CDR (STCDR) by utilizing data from the source domains to improve the model's performance on the target domain, or applied dual-target CDR (DTCDR) by integrating data from the source and target domains. In addition, multi-target CDR (MTCDR) is a generalization of DTCDR, which is able to capture the link among different domains. In this paper we present HGDR (Heterogeneous Graph-based Framework with Disentangled Representations Learning), an end-to-end heterogeneous network architecture where graph convolutional layers are applied to model relations among different domains, meanwhile utilizes the idea of disentangling representation for domain-shared and domain-specifc information. First, a shared heterogeneous graph is generated by gathering users and items from several domains without any further side information. Second, we use HGDR to compute disentangled representations for users and items in all domains. Experiments on real-world datasets and online $\mathrm{A} / \mathrm{B}$ tests prove that our proposed model can transmit information among domains effectively and reach the SOTA performance. The code can be found here: https://github.com/NetEase-Media/HGCDR. 
\end{abstract}



\maketitle

\section*{1 INTRODUCTION}
Recommender systems are playing more and more important roles in online personalized services. As an important tool to understand user intentions, recommender systems focus on discovering the most relevant information for their consumers to increase user engagement and reduce the time spent searching for it. However, even through they can capture different latent relationships between users and items, they are still suffering from the sparsity issue.

Recently, Cross-Domain Recommendation (CDR) \cite{ref12} is a potential strategy to mitigate the sparsity issue. It integrates large amounts of data from one domain to another to improve recommendation performance. Previous research on Single-target CDR (STCDR) and Dual-target CDR (DTCDR) is relatively sufficient, but there is lack of attention on the Multi-target CDR (MTCDR), a generalization of DTCDR whose purpose is to improve recommendation performance of multiple domains. MTCDR faces a lot of challenges in the real-world recommendation system, including model complexity and difficulty of improvement on multiple domains.

In recent related research, graph neural networks (GNNs) have shown great potential in the field of recommendation algorithms \cite{ref15}. Unlike traditional deep models, GNNs can model non-Euclidean data with a variable number of neighbours. In the field of recommendation algorithms, we are generally dealing with heterogeneous graphs with several types of nodes and edges. For example, in an ecommerce scenario, nodes can be "items", "shops", "users" and edge types can be "click", "favorite". Relationship Graph Convolutional Network (RGCN) \cite{ref11} is the first GNN model to consider heterogeneity, modelling the heterogeneity of edges through edgetypespecific transformation.

In this paper, we represent users and items from different domains as nodes of different types in the graph, and their interactions as edges of different types in the global graph if the user interacts with the item in different domains. Without side information, we use the IDs of the nodes as the only input. Different from previous GNN based models, our proposed model utilizes the idea of disentangling the domain-shared and domain-specifc information, which has been proved to be an efficient way to transfer information across different domains. This study makes the following major contributions: (1) We propose a new graph convolutional layer designed for MTCDR, which aggregates neighbor nodes while considering types of relationship and nodes' domains; (2) We design an efficient way to learn disentangled representations for domain-shared and domain-specifc information for users and items in different domains; (3) Experiments on the real-world datasets show that our proposed model achieves significantly better performance than other state-of-the-art models and online A/B tests further prove that our proposed model achieves better performance than many online models.

\section*{2 RELATED WORK}
\subsection*{2.1 Cross Domain Recommendation}
The main feature of cross-domain recommendation is that it can be used to capture user preferences in certain areas by analyzing information about their interactions in other domains. This can then be used to enrich the data in the target domain. There are two types of cross-domain recommendations: (1) Asymmetric Approach: it uses data in the source domain to address data sparsity in the target domain, specifically by applying knowledge learned in the source domain or some pattern directly to the target domain acting as prior knowledge. (2) Symmetric Approach: it assumes that both the source and target domains have a data sparsity issue and they are treated equally. Typically this approach learns a mapping function between the domains that makes a clear distinction between factors that are unique to each domains and factors that are shared between domains. Many different kinds of cross-domain recommendation methods have been proposed, including DTCDR (Dual-Target CrossDomain Recommendation) \cite{ref18}, EMCDR (Embedding and Mapping Cross-Domain Recommendation) \cite{ref7}, CoNet(Collaborative Cross Networks) \cite{ref4}, HeroGRAPH \cite{ref9} etc.

\subsection*{2.2 Graph neural networks for recommendation}
There is only one type of node and edge in the homogeneous graph, so all nodes share the same model parameters and have the same dimensional feature space when building a graph neural network In contrast, there can be more than one type of node and edge in a heterogeneous graph, allowing different types of nodes to have different dimensions of features or attributes. Common recommen dation system data consists of information about the interactions between users and items. The graph is a heterogeneous graph since user nodes may contain personal information such as the user's age and occupation, while item nodes contain item-specific information such as content and price. Many different kinds of models with graph neural networks have been proposed, including GCMC (Graph Convolutional Matrix Completion) \cite{ref1}, ACKRec (Attention Concept Knowledge Recommendation) \cite{ref14}, GHCF (Graph Heterogeneous Collaborative Filtering) \cite{ref6}, etc.

\subsection*{2.3 Disentangled Representation Learning}
Disentangled representation learning focuses on factorizing the latent factors from data, which achieves great success in many areas such as computer vision. Disentangled representation learning in recommendation is unexplored until recently, several models have been proposed, including DICE (Disentangling Interest and Conformity with Causal Embedding) \cite{ref17}, CLSR (Contrastive learning framework to disentangle Long and Short-term interests for Rec ommendation) \cite{ref16}, DisenCDR (Disentangled Representations for Cross-Domain Recommendation) \cite{ref5}, etc. DisenCDR utilizes two mutual-information-based regularizers to disentangle the domainshared and domain-specifc information, which inspired our present graph-based method.

\section*{3 METHODOLOGY}
In this section, firstly we formalize the multi-target cross-domain recommendation issue by introducing one heterogeneous graph.\\
Then we propose our novel heterogeneous graph network HGDR (Heterogeneous Graph-based Framework with Disentangled Representations Learning). Finally, we describe how to use BPR loss function to train our proposed model.

\subsection*{3.1 Preliminary Concepts and Notations}
Let $U$ and $I$ stand for user and item sets, $D$ stands for the domain set and domain $d$ is represented by the subscript $d$. We only use the user ID, item ID, and their interactions to build one global heterogeneous graph. We define the interaction as edges that connect users and items. Since our data is modeled on a heterogeneous graph, our edges also have different types. For example, for edges between one user and items from different domains, "\textit{user - click - music}" and "\textit{user - click - movie}" are two types of edges, similarily, "\textit{user - click - music}" and "\textit{music - click - by - user}" are different types of edges.

\subsection*{3.2 Model Structure}
\begin{figure*}[h]
  \centering
  \includegraphics[width=\textwidth]{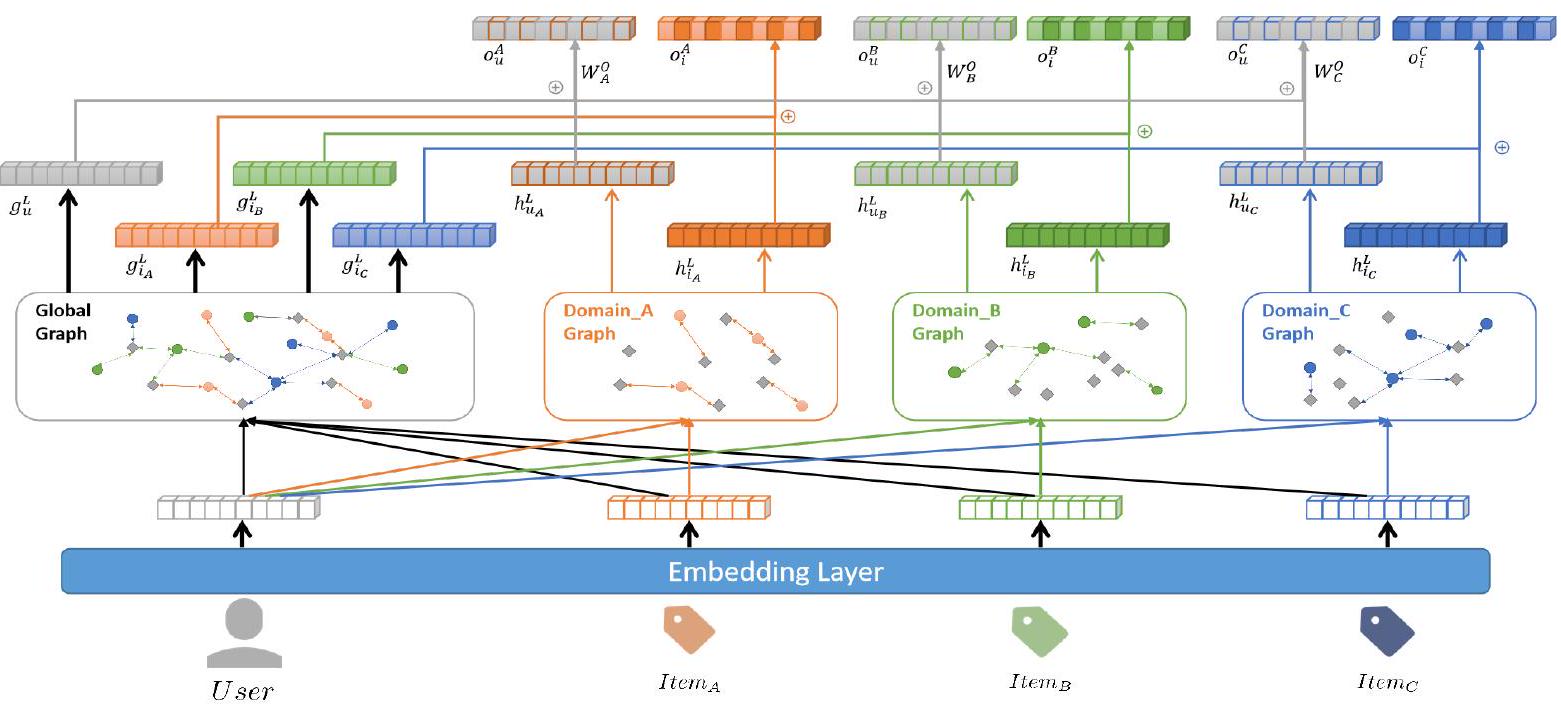}
  \caption{Our proposed HGDR framework}
  \label{fig:figure 1}
\end{figure*}

HGDR consists of two main parts, firstly a mapping layer structure that generates embedding based on node IDs, and a newly proposed disentangled convolutional structure including domain-specific graph convolution and domain-shared graph convolution, which is used to learn disentangled representations for nodes in different domains. The final output for a node is the integration of the outputs of the above two graph convolution structures. The overall structure of our proposed model is depicted on Figure \ref{fig:figure 1}.

3.2.1 \textit{Mapping Layer}. We generate embeddings for each user and item in one common semantic context. We only have IDs for users and items from different domains because there is no side information, thus we can assign an embedding for each ID, which is represented as ( $\left.e_{u}, e_{i}, e_{j}, \ldots\right)$, where $e_{u}$ represents the embedding for user entities and $e_{i}, e_{j}$ represents embeddings for item entities. We follow the general scenario that users are shared in all domains, while each item belongs to only one domain.

3.2.2 \textit{Disentangled Convolutional Layer}. After obtaining embeddings for users and items according to their IDs, we use one heterogeneous graph to connect users and items from different domains. Inspired by RGCN and DisenCDR, we design the following domainspecific graph convolution operation and domain-shared graph convolution operation to learn disentangled representations for nodes in different domains.

\begin{itemize}
  \item \textbf{domain-specific graph convolution:}
\end{itemize}

In order to model different user preferences of different domains and mitigate the transferring of useless information among different domains, for each node in domian $d$, the domain-specific graph convolution operation aggregates information of neighbours in the same domain. More specifically, for user $u$, the domain-specific representation of $(l+1)$ th layer is get by:

\begin{equation*}
h_{u_{d}}^{l+1}=\operatorname{Relu}\left(h_{u_{d}}^{l} W_{d_{U U}}^{l}+\sum_{i \in N_{u}^{d}} h_{i_{d}}^{l} W_{d_{I U}}^{l}\right) \tag{1}
\end{equation*}

and for item $i$

\begin{equation*}
h_{i_{d}}^{l+1}=\operatorname{Relu}\left(h_{i_{d}}^{l} W_{d_{I I}}^{l}+\sum_{u \in N_{i}^{d}} h_{u_{d}}^{l} W_{d_{U I}}^{l}\right) \tag{2}
\end{equation*}

where $W_{d_{(\cdot)}}^{l}$ represents the learnable feature transform matrix for different relations, here we consider four different types including "user to user"(UU), "item to user"(IU), "item to item"(II) and "item to user"(IU), $N_{u}^{d}$ and $N_{i}^{d}$ represents the neighbourhood nodes of user and item in domain $d, h_{u_{d}}^{0}=e_{u}, h_{i_{d}}^{0}=e_{i}$, and Relu represents the activation function;

\begin{itemize}
  \item \textbf{domain-shared graph convolution:}
\end{itemize}

In order to get valuable information for transfering among different domains, we design the domain-shared graph convolution operation to aggregate information of neighbours in all domains. More specifically, for user $u$, the domain-shared representation of $(l+1)$ th layer is get by:

\begin{equation*}
g_{u}^{l+1}=\operatorname{Relu}\left(g_{u}^{l} W_{s U U}^{l}+\sum_{d \in D} \sum_{i \in N_{u}^{d}} g_{i}^{l} W_{d_{I U}}^{l}\right) \tag{3}
\end{equation*}

and for item $i$ :

\begin{equation*}
g_{i}^{l+1}=\operatorname{Relu}\left(g_{i}^{l} W_{s_{I I}}^{l}+\sum_{d \in D} \sum_{u \in N_{i}^{d}} g_{u}^{l} W_{d U I}^{l}\right) \tag{4}
\end{equation*}

where $W_{s_{(.)}}^{l}$ represents the learnable feature transform matrix for shared graph convolution, $g_{u}^{0}=e_{u}, g_{i}^{0}=e_{i}$. It is worthwhile to notice that many attention-based graph convolution operations such as GAT (Graph Attention Network) can be directly applied in our proposed models.

3.2.3 \textit{Output Layer}. The final representation of each node is integrated by the domain-specific representation and domain-shared representation:

\begin{equation*}
o_{u}^{d}=\left(h_{u_{d}}^{L}+g_{u}^{L}\right) W_{d}^{O}, \quad o_{i}^{d}=\left(h_{i_{d}}^{L}+g_{i}^{L}\right) \tag{5}
\end{equation*}

where $L$ is the last convolutional layer and $W_{d}^{O}$ is learnable output transform parameters for users.

\subsection*{3.3 Training and Optimization Process}
In this subsection, we calculate the user preference and train the model. The positive user preference is calculated based on the matrix factorization:

\begin{equation*}
x_{u i}^{d}=o_{u}^{d} o_{i}^{d} \tag{6}
\end{equation*}

Then we apply the widely-used pair-wise Bayesian Personalized Ranking (BPR) to train the model:

\begin{equation*}
l_{u i j}^{d}=-\ln \sigma\left(x_{u i}^{d}-x_{u j}^{d}\right) \tag{7}
\end{equation*}

where $x_{u j}^{d}=o_{u}^{d} o_{j}^{d}$ is negative preference based on negative feedback pair $\left(u_{d}, j_{d}\right)$. Finally the loss function for domain $d$ is:

\begin{equation*}
\theta_{d}^{*}=\underset{\theta}{\operatorname{argmin}} \sum_{u} l_{u i j}^{d}+\lambda_{\theta}\|\theta\|^{2} \tag{8}
\end{equation*}

The total loss is the sum of each domain's loss: $\theta^{*}=\sum_{d \in D} \beta_{d} \theta_{d}^{*}$, where $\beta_{d}$ are hyper-parameters set according to the data size of each domain, all parameters are updated by Adam \cite{ref2} with default values.

\section*{4 EXPERIMENTS}
PyTorch \cite{ref8} and the DGL \cite{ref13} are used to implement our proposed model. The model has two disentangled convolutional layers and the embedding dimension is 128 .

\subsection*{4.1 Datasets and Evaluation Protocols}
\textbf{Datasets}: Generally we follow experiment settings in GA-DTCDR \cite{ref3}. Two experiments are carried out using the Douban dataset, and our industrial dataset which is collected from three recommendation scenarios of our APP in one day. Table \ref{tab:tab 1} shows the statistics of the two datasets.

\begin{table}[h]
  \caption{Statistics of the two datasets}
  \label{tab:tab 1}
  \renewcommand{\arraystretch}{1.1} 
  \begin{tabular}{c|ccc}
  \toprule
  Dataset & \multicolumn{3}{c}{Douban}  \\
  \midrule
  Domain & Book & Music & Movie  \\
  \midrule
  \# Users & 2110 & 1672 & 2712  \\
  \# Items & 6777 & 5567 & 34893  \\
  \# Interactions & 96041 & 69709 & 1278401  \\
  \midrule
  sparsity (\%) & 0.67 & 0.75 & 1.35  \\
  \bottomrule
  \toprule
  Dataset & \multicolumn{3}{c}{Our App}  \\
  \midrule
  Domain & Scenario A & Scenario B & Scenario C  \\
  \midrule
  \# Users & 88491 & 77609 & 172076  \\
  \# Items & 193521 & 211236 & 114550  \\
  \# Interactions & 2417083 & 1361152 & 696868  \\
  \midrule
  sparsity (\%) & 0.0141 & 0.0083 & 0.0035  \\
  \bottomrule
  \end{tabular}
\end{table}

\textbf{Evaluation Protocols}: For each test user, we choose the latest interaction with a test item as the test interaction and randomly sample 99 unobserved interactions for the test user, and then rank the test item among the 100 items. HR@10 (Hit Rate) and NDCG@10 (Normalized Discounted Cumulative Gain) are used as the performance metrics.

\subsection*{4.2 Benchmarks}
We choose several methods to compare the proposed model:

(1) \textbf{MF (Matrix Factorization)} \cite{ref10}: MF is one powerful single domain recommendation method, which uses implicit feedback from users to rank items by the maximum posterior probability obtained from a Bayesian analysis of the problem.

(2) \textbf{DDTCDR (Dual-Target Cross-Domain Recommendation)} \cite{ref18}: DDTCDR is an effective cross-domain recommendation framework based on multi-task learning. Given two domains containing user and item ratings, textual information, etc., the model investigates the goal of recommending more accurate items to users in both domains simultaneously.

(3) \textbf{GA-DTCDR (Graphical and Attentional framework for Dual-Target Cross-Domain Recommendation)} \cite{ref3}: GA-DTCDR is a unified framework for dual-target CDR and multi-target CDR based on Graph embedding and Attention techniques, in which the element-wise attention mechanism and the personalized training strategy are used to improve accuracy and avoid negative transfer

(4) \textbf{RGCN (Relational Graph Convolutional Network)} \cite{ref11} RGCN is the first paper in graph neural network field that considers edge difference. It utilizes different propagation layers for different types of edge to capture edge information, which are naturally suited for cross-domain recommendation.

\subsection*{4.3 Results}
We first compare our method with prior SOTA methods in the two experiments. The results are shown in Table \ref{tab:tab 2}. The results show that our proposed model achieves the best performance among all methods.

\begin{table}[h]
  \caption{Performance Comparison}
  \label{tab:tab 2}
  \renewcommand{\arraystretch}{1.1} 
  \begin{tabular}{c|ccc}
  \toprule
  Experiment & \multicolumn{3}{c}{Exp. 1 (Douban)} \\
  \midrule
  Domain & Book & Music & Movie \\
  \midrule
  Metric & \multicolumn{3}{c}{HR(\%) / NDCG(\%)} \\
  \midrule
  MF & $0.577 / 0.351$ & $0.330 / 0.197$ & $0.404 / 0.245$ \\
  DDTCDR & $0.586 / 0.358$ & $0.352 / 0.211$ & $0.422 / 0.243$ \\
  RGCN & $0.697 / 0.438$ & $0.420 / 0.235$ & $0.474 / 0.287$ \\
  GA-DTCDR & $0.695 / 0.441$ & $0.449 / 0.260$ & $0.499 / 0.309$ \\
  \midrule
  HGDR(Our) & $\textbf{0.712} / \textbf{0.448}$ & $\textbf{0.483} / \textbf{0.281}$ & $\textbf{0.550} / \textbf{0.346}$ \\
  \bottomrule
  \toprule
  Experiment & \multicolumn{3}{c}{Exp. 2 (Our App)} \\
  \midrule
  Domain & Scenario A & Scenario B & Scenario C \\
  \midrule
  Metric & \multicolumn{3}{c}{HR(\%) / NDCG(\%)} \\
  \midrule
  $\mathrm{MF}$ & $0.577 /0.411$& $0.591 / 0.433$ & $0.322 / 0.203$\\
  $\mathrm{RGCN}$ & $0.697 / 0.513$ & $0.665 / 0.503$ & $0.751 / 0.613$ \\
  \midrule
  HGDR(Our) & $\textbf{0.721} / \textbf{0.534}$ & $\textbf{0.692} / \textbf{0.529}$ & $\textbf{0.775} / \textbf{0.629}$ \\
  \bottomrule
  \end{tabular}
\end{table}

\subsection*{4.4 Online Evaluation}
To verify the effectiveness of HGDR in real-world scenarios, we conduct an online $\mathrm{A} / \mathrm{B}$ test on our online recommendation system of Our APP. Precisely, we deploy HGDR and several competitive baselines in the matching module of a cross-domain sceniaro, with the ranking module unchanged. In online evaluation, we focus on three online metrics in the target domain: (1) CTR, (2) average user duration (seconds), and (3) coverage ratio for users. We conduct the A/B test for 5 days, and results are shown in Table \ref{tab:tab 3}

\begin{table}[h]
  \caption{Online A/B tests on Our App}
  \label{tab:tab 3}
  \renewcommand{\arraystretch}{1.1} 
  \label{tab:freq}
  \begin{tabular}{c|ccc}
  \toprule
  & CTR(\%) & Duration(s) & Coverage Ratio(\%) \\
  \midrule
  MF & 6.12 & 85.69 & 32.28 \\
  RGCN & 7.48 & 90.78 & 32.31 \\
  \midrule
  HGDR(Our) & $\mathbf{\textbf{8.35}}$ & $\mathbf{\textbf{94.78}}$ & $\mathbf{\textbf{34.1}}$ \\
  \bottomrule
  \end{tabular}
\end{table}

\section*{5 CONCLUSIONS AND FUTURE WORK}
In this work, we propose a novel heterogeneous graph based framework with disentangled representations learning structure, which is used for learning domain-shared and domain-specifc information for cross-domain recommendation. Experiments on the real-world datasets prove that our proposed model achieves much better performance than other state-of-the-art models, and online A/B tests further prove that our proposed model achieves better performance than many online models. In the future, we will explore the possibility of further improving our models with users' or items' side information. Furthermore, we will try to utilize multi-modal information as input to improve our model performance in the field of cross domain recommendation.

\end{document}